\documentclass[epj]{webofc}
\usepackage[varg]{txfonts}   
\wocname{EPJ Web of Conferences}
\woctitle{CONF12}
\newcommand{\conds}{\langle \bar s s \rangle}

\newcommand{\quarkcorT}{\langle {\cal T} (\bar \psi_l\psi_l)(x) \,(\bar \psi_l
 \psi_l)(0)\rangle_{T}}
\newcommand{\Od}{{\cal O}}
\newcommand{\mean}[1]{\left\langle{#1}\right\rangle}
\newcommand{\condl}{\mean{\bar \psi_l \psi_l}}
\newcommand{\im}{\mbox{Im}\,}
\newcommand{\re}{\mbox{Re}\,}

\begin{document}
\selectlanguage{english}
\title{Chiral Symmetry restoration from the hadronic regime}
%
%

\author{Angel G\'omez Nicola\inst{1}\fnsep\thanks{\email{gomez@ucm.es}} \and
        Santiago Cort\'es\inst{2} \and
        John Morales\inst{3} \and
       Jacobo Ruiz de Elvira \inst{4,5}
       \and
       Ricardo Torres Andr\'es\inst{6}
}

\institute{Departamento de F\'{\i}sica
Te\'orica II. Univ. Complutense. 28040 Madrid. Spain
\and
           Departamento de F\'{\i}sica, Univ. de Los  Andes, 111711 Bogot\'a, Colombia
\and
          Departamento de F\'{\i}sica, Univ. Nacional de Colombia, 111321 Bogot\'a, Colombia
           \and
           Albert Einstein Center for Fundamental Physics, Institute for Theoretical Physics, University of Bern, Sidlerstrasse 5, CH--3012 Bern, Switzerland 
           \and
          Helmholtz--Institut f\"ur Strahlen- und Kernphysik (Theorie) and   Bethe Center for Theoretical Physics, Universit\"at Bonn, D--53115 Bonn, Germany
           \and
IES Gerardo Diego, 28224 Pozuelo de Alarc\'on (Madrid), Spain
}

\abstract{%
  We discuss recent advances on QCD chiral symmetry restoration at finite temperature, within the theoretical framework of Effective Theories.   $U(3)$ Ward Identities are derived between pseudoscalar susceptibilities and quark condensates, allowing to explain the behaviour of lattice meson screening masses. Unitarized interactions and the generated $f_0(500)$ thermal state are showed to play an essential role in the description of the transition through the scalar susceptibility. 
}
\maketitle
\section{Introduction}
\label{intro}
Chiral symmetry restoration is a fundamental part of the QCD phase diagram and  plays an essential role in the understanding of the Physics of strong interactions in extreme conditions, such as those reached in Relativistic Heavy Ion Collisions. It  is important to provide solid theoretical analysis regarding  chiral restoration, given the limitations of perturbative QCD at those temperature scales. We review here recent advances on the description of  such environment. Most of our theoretical work has been carried out within the Chiral Effective Lagrangian framework. For the relevant temperature range, one needs an effective description accounting for the degrees of freedom involved, which at low and moderate temperatures below the transition $T_c$ are just pions, the more abundant ones. Thus, our main theoretical setup to describe the physics involved has been Chiral Perturbation Theory (ChPT) \cite{Gerber:1988tt} and its unitarized versions (see  Sects.~\ref{unit} and ~\ref{largeN}). Within that framework several properties of interest for the hadron gas have been studied. In particular,  unitarized interactions have allowed to describe properly thermal resonances \cite{Dobado:2002xf} and transport coefficients \cite{FernandezFraile:2009mi}.  Here we will first discuss general ideas about partners and patterns (Sect.~\ref{part}) in the context of lattice and effective theory analysis of susceptibilities in the scalar and pseudoscalar sectors \cite{Nicola:2013vma}. In  Sect.~\ref{Ward}, we will show that QCD Ward identities relating pseudoscalar susceptibilities and quark condensates allow to describe the behaviour of meson screening masses near chiral restoration \cite{Nicola:2013vma,Nicola:2016jlj}.  Sect.~\ref{unit} describes the role of the thermal $f_0(500)$ state  for the description of the scalar susceptibility \cite{Nicola:2013vma}. Finally, recent analysis  within the large number of Nambu-Goldstone Bosons (NGB) framework  \cite{Cortes:2015emo,Cortes:2016ecy}  will be presented in Sect.~\ref{largeN}.  
\section{Chiral partners and patterns}
\label{part}

The main properties of the chiral phase transition at finite temperature $T$ and zero baryon density have been established recently by lattice simulations. For $N_f=2+1$ flavours and physical quark masses, the transition is a crossover at $T_c\sim$ 155 MeV,  signaled by the peak of the  scalar susceptibility and the inflection point of the asymptotically vanishing light quark condensate $\condl$ \cite{Aoki:2009sc,Bazavov:2011nk}.  However there are still open problems, like the behavior under different degeneracy patterns and the corresponding chiral partners. For instance, the chiral partner of the pion for a given  pattern. Thus, the early $O(4)\rightarrow O(3)$ pattern for chiral symmetry breaking \cite{Pisarski:1983ms} would imply degeneration of $\pi-\sigma$ states, esentially by $\sigma$ mass dropping driven by the  order parameter $\mean{\sigma}\sim \condl$. However, it is nowadays well established that the $\sigma$ state with isospin and total angular momentum $I=J=0$ is actually a broad resonance  in $\pi\pi$ scattering, denoted $f_0(500)$ in the modern notation \cite{Pelaez:2015qba}.  A consistent way to describe those chiral partners without assuming any particular nature for the $f_0(500)$ can be achieved by studying the corresponding scalar and pseudoscalar susceptibilities, which for the light sector read:
\begin{eqnarray}
\chi_S^\sigma (T)&=&-\frac{\partial}{\partial m_l} \condl(T)=\int_E{d^4x \left[\quarkcorT-\condl^2\right]}\nonumber\\&=&\frac{1}{Z}\int_E{d^4x \left[\frac{\delta}{\delta s(x)}\frac{\delta}{\delta s(0)}Z[s,p]\bigg\vert_{s={\cal M},p^a=0}\right]},\label{chisdef}\\
 \chi_P ^\pi(T)\delta^{ab}&=&\int_E{d^4x \langle {\cal T} \pi^a (x) \pi^b (0)  \rangle_T}= \frac{1}{Z}\int_E{d^4x \left[\frac{\delta}{\delta p^a(x)}\frac{\delta}{\delta p^b(0)}Z[s,p]\bigg\vert_{s={\cal M},p^a=0}\right]},\label{chipdef}
\end{eqnarray}
where ${\cal M}=\mbox{diag}(m_l,m_l,m_s)$ is the quark mass matrix, $\int_E$ extends over euclidean space-time, $Z=Z[s={\cal M},p=0]$ is the QCD partition function and $s(x), p(x)$ are scalar and pseudoscalar sources.  The quark bilinears entering have the quantum numbers of the pion and $f_0(500)$ states:
\begin{equation}
\pi^a=\bar\psi_l\gamma_5\tau^a\psi_l; \quad 
\sigma=\bar\psi_l \psi_l.
\label{pisigmabi}
\end{equation}

\begin{figure}[h]
\centering
\includegraphics[width=7cm,clip]{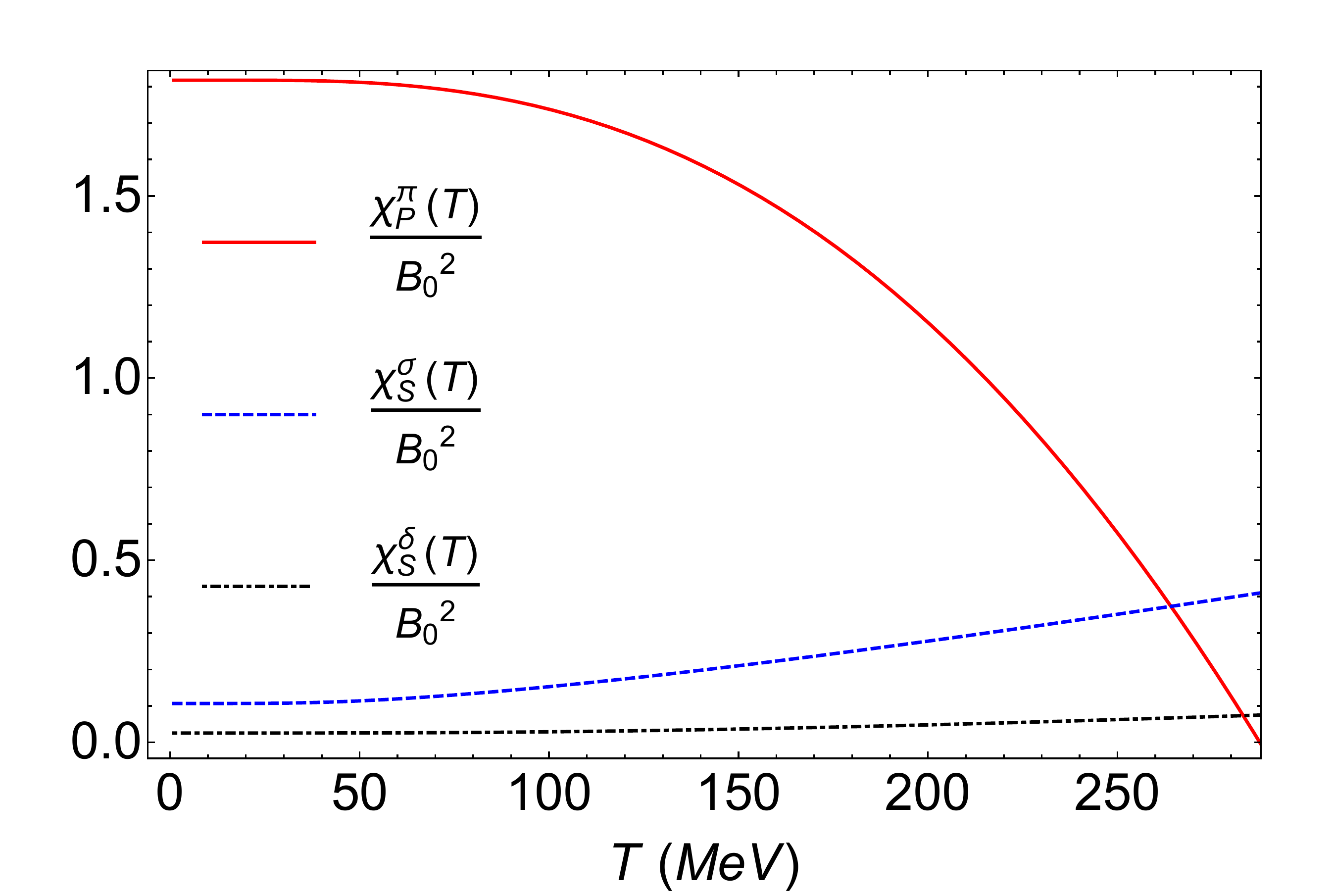}
\includegraphics[width=7cm,clip]{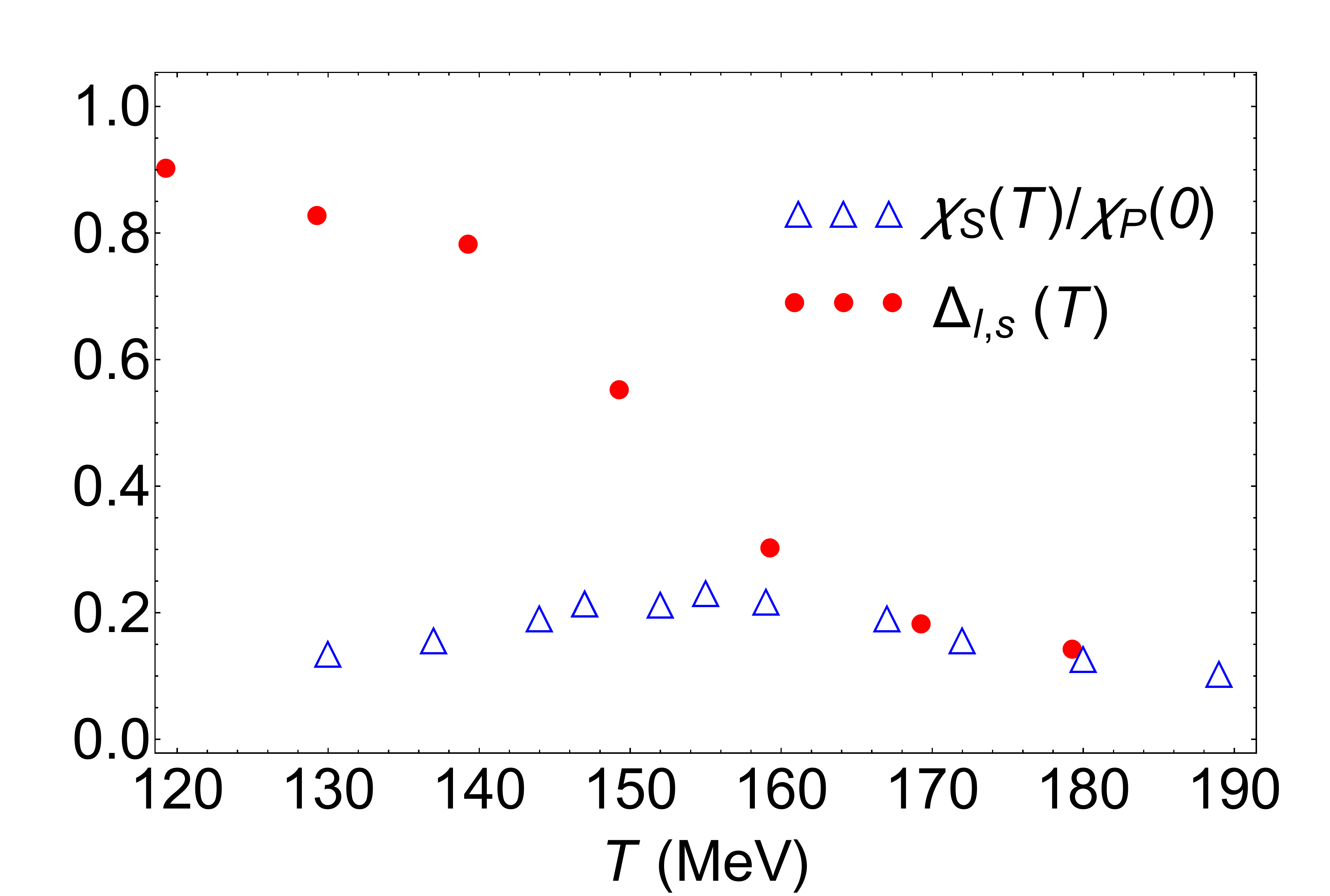}
\caption{Left: $U(3)$ ChPT one-loop results for susceptibilities. Right: Scalar-pseudoscalar degeneration in the $\pi-\sigma$ sector from lattice results in \cite{Aoki:2009sc}}
\label{fig-scalarpseudolatt}       
\end{figure}

Therefore, within the $O(4)$ pattern, $\chi_S(T)$ and $\chi_P(T)$ should become degenerate near the maximum of $\chi_S(T)$. This is actually the case in the lattice, as can be seen from figure~\ref{fig-scalarpseudolatt}, where we show lattice data from \cite{Aoki:2009sc} for $\chi_S$ and the ratio of subtracted quark condensates $\Delta_{l,s}$, which through the Ward identities explained in Sect.~\ref{Ward}, has the same temperature dependence as $\chi_P(T)/\chi_P(0)$ \cite{Nicola:2013vma}. Direct comparison between $\chi_S$ and $\chi_P$ in \cite{Buchoff:2013nra} shows  the same degeneration. On the other hand,  ChPT  to one loop  yields an increasing $\chi_S(T)$, intersecting   $\chi_P(T)$ at $T_d\simeq 0.9 T_c$, where $\condl^{ChPT} (T_c)=0$. Although this is  just  an extrapolation of the model-independent expressions for $\chi_S(T)$ and $\chi_P(T)$ near $T_c$,  ChPT supports the idea of partner degeneration. Actually, near the chiral limit $M_\pi\ll T$, where  critical effects are meant to be enhanced and different critical points should coincide in a true phase transition, both temperatures degenerate as  $T_d=T_c-3M_\pi/4\pi+\Od(M_\pi^2/T_c)$. 

An important related question is whether the $U_A(1)$ symmetry is restored or not at the chiral transition. This is still a matter of debate in lattice analysis and is usually addressed by examining, in addition to the states in eq.\eqref{pisigmabi}, 
\begin{equation}
\delta^a=\bar\psi_l \tau^a \psi_l\quad  \eta_l=\bar\psi_l \gamma_5 \psi_l,
\end{equation}
which correspond physically to the $a_0(980)$ ($\delta$) and the light part of the $\eta$. Thus, a pure $SU_A(2)$ transformation allows to connect  $\pi^a-\sigma$  (as in the $O(4)$ pattern) and $\delta^a-\eta_l$ states while a $U_A(1)$ connects $\pi^a-\delta^a$ and $\eta_l-\sigma$. Although the $U_A(1)$ symmetry is broken by the axial anomaly, it could in principle be restored at high enough temperature, and if that was the case, degeneration of the above patterns should be observed in lattice simulations around $T_c$. The symmetry breaking pattern would be then $O(4)\times U_A(1)$, which would mean in particular that the $\eta'$  plays an active role since it becomes a ninth Goldstone boson, as in the large-$N_c$ context  \cite{'tHooft:1973jz}.  Actually, the ChPT framework can be  extended by adding the $1/N_c$ counting to the standard chiral low-energy one, describing consistently the $\eta'$ and the $U_A(1)$ anomaly 
\cite{DiVecchia:1980ve}, giving rise to the so called $U(3)$ ChPT.

Regarding the lattice results for $U_A(1)$ restoration,    the susceptibilities associated to the above quark bilinears have been studied in \cite{Buchoff:2013nra} and it has been found that 
$\chi_\pi-\chi_\delta$ and $\chi_\sigma-\chi_{\eta_l}$ vanish  asymptotically  well above $T_c$, so  the transition pattern remains $O(4)$, consistent also with earlier analysis of the same group on screening masses \cite{Cheng:2010fe}.  However, the correlator analysis of \cite{Cossu:2013uua} is   compatible with $\pi-\delta-\sigma-\eta_l$ degeneration and thus with $U_A(1)$ restoration at the chiral transition.  In addition,  recent results in \cite{Brandt:2016daq} show that the difference of  screening masses of the $\pi$ and $\delta$ channel are found to be  compatible with zero at the chiral transition.

In figure~\ref{fig-scalarpseudolatt} we show the results of $U(3)$ ChPT for the above susceptibilities, calculated to one loop and with standard typical values for the low-energy constants involved, where $B_0=M_\pi^2/2m_l$ to leading order. We observe that $\chi_\delta$ matches $\chi_\pi$ above the point where $\chi_\sigma-\chi_\pi$ match, corresponding to a linear growth  $\chi_S^\sigma\propto T/M_\pi$ (a similar growing behaviour is obtained in the virial approach \cite{GomezNicola:2012uc}) and a quadratic  one $\chi_S^\delta \sim \propto T^2/M_\eta^2$ suppressed by the inverse strange mass \cite{Nicola:2011gq}. One could then argue that ChPT favors $U_A(1)$ restoration at a higher temperature than  chiral $O(4)$. However, we must keep in mind that the gap between $O(4)$ $\chi_S^\sigma-\chi_P^\pi$ intersection and the vanishing of the quark condensate is relatively small and actually tends to vanish in the chiral limit, as discussed above. Therefore,   ChPT would also be compatible with the $O(4)\times U_A(1)$ pattern from this viewpoint of partner degeneration. In addition, the temperatures involved are beyond the strict ChPT validity range, which adds more uncertainty  to this conclusion. In this sense, note that  the critical  temperatures in ChPT are higher than  lattice values, which is a known feature of a pure NGB light meson description, to be improved by considering higher orders and heavier hadron states  \cite{Gerber:1988tt,Huovinen:2009yb}. 
\section{Ward Identities and lattice screening masses}
\label{Ward}

As it is clear from the discussion in the previous section, scalar and pseudoscalar susceptibilities play a crucial role to understand the nature of the chiral phase transition. In the  pseudoscalar case, we have seen that $\chi_\pi$ actually inherits the critical properties through its direct relation with the light quark condensate. That relation is  a particular example of a set of Ward Identities (WI) obtained formally from QCD by performing $U(3)$ axial transformations, extending previous works \cite{Bochicchio:1985xa},   and verified explicitly in $U(3)$ ChPT \cite{Nicola:2016jlj}. For the $\pi$, $K$ and $\bar s s$ sector, they read:
\begin{eqnarray}
m_l \,\chi_P^\pi(T)&=&-\condl  (T)
\label{wipi}\\
\left(m_l+m_s\right) \chi_P^K(T)&=&-\left[\condl (T)+ 2 \left\langle\bar s s \right\rangle (T)\right]
\label{wik}\\
\chi_P^{\bar s s}&=&-\frac{\mean{\bar ss}}{m_s}+\frac{m_l}{4\sqrt{3}  m_s\left(m_l -m_s\right)}\chi_P^{8A},
\label{wis}
\end{eqnarray}
where the above pseudoscalar susceptibilities are constructed analogously to the $\pi$ channel  in \eqref{chipdef}, and  
$\chi_P^{8A}$ corresponds  to a  crossed correlator between $\bar\psi\lambda^8\psi$ and the anomalous operator
$A(x)=\frac{3g^2}{32\pi^2}G_{\mu\nu}^a\tilde G^{\mu\nu}_a$. Note that in the $U(3)$ case, the octet member $\eta_8$, the singlet $\eta_0$ and the anomaly $A$ are mixed. Recall also that in the $\bar s s$ channel equation \eqref{wis}, the anomalous term is suppressed by a $m_l/m_s$ factor with respect to the nonanomalous one. That suppression is confirmed by the lattice direct check of this identity in \cite{Buchoff:2013nra}, where the $\pi$-channel one is also checked. However, to the best of our knowledge, there is no direct check of the $K$ channel.

\begin{figure}[h]
\centering
\includegraphics[width=7cm,clip]{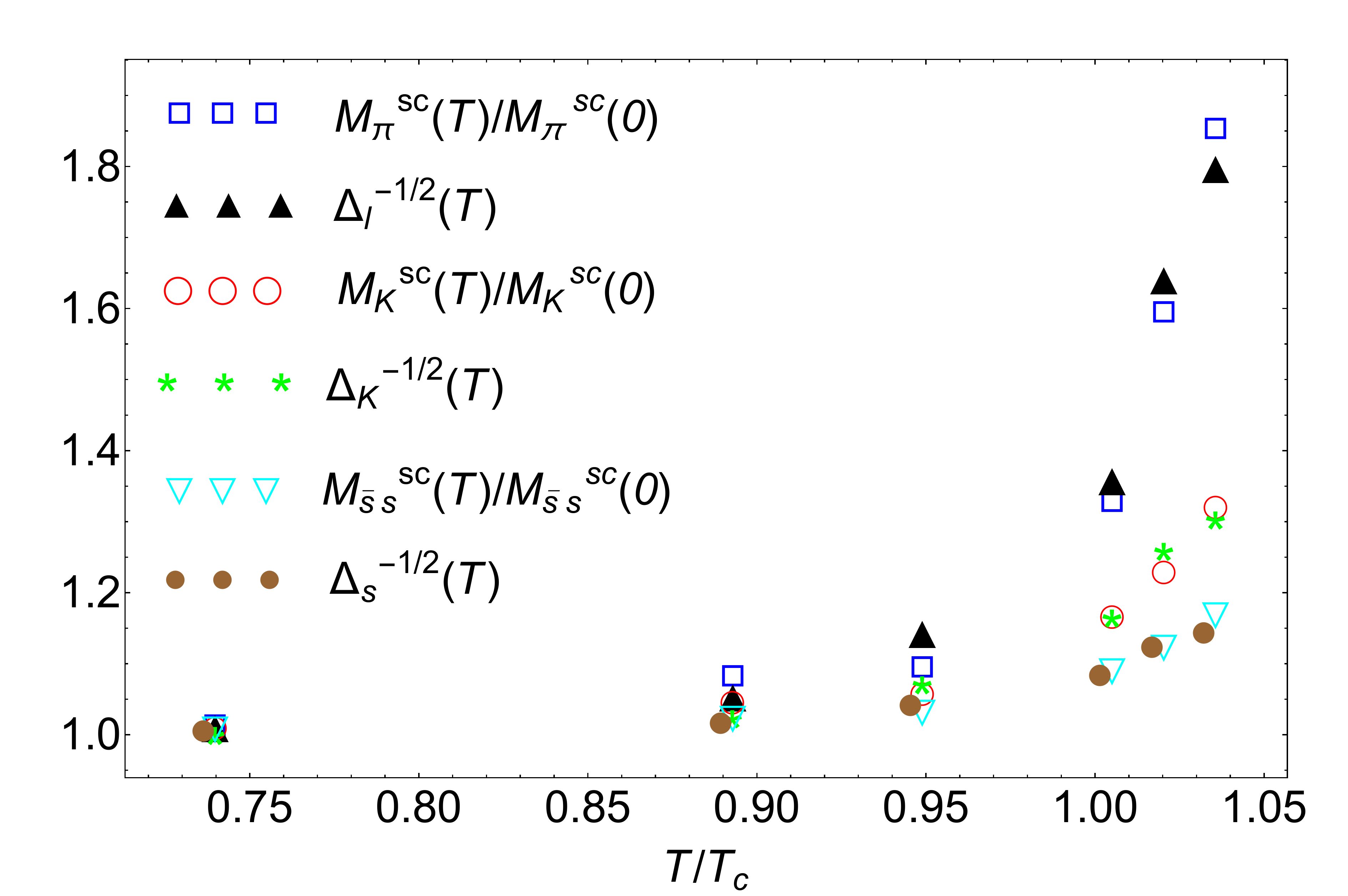}
\caption{Comparison between screening masses and subtracted quark condensates, according to the WI scaling \eqref{scalingpi}, \eqref{scalingK}, \eqref{scalings}.}
\label{fig-screening}       
\end{figure}

The above WI turn out to be a powerful tool to explain the behaviour of   meson screening masses $M_i^{sc}$ for those channels,  measured in the lattice at finite temperature.  Under certain assumptions, they can be related to susceptibilities \cite{Nicola:2016jlj}. Actually, the latter  are nothing but the zero momentum limit of correlators so that we expect $\chi\sim M^{-2}$ with $M$ the mass of the corresponding channel. In particular, from \eqref{wipi}, \eqref{wik}, \eqref{wis}, one is led to the following scaling behaviour \cite{Nicola:2016jlj}:
\begin{align}
\frac{M^{sc}_\pi (T)}{M^{sc}_\pi (0)} \sim \left[\frac{\chi_P^\pi(0)}{\chi_P^\pi(T)}\right]^{1/2}=\left[\frac{\condl(0)}{\condl (T)}\right]^{1/2}\label{scalingpi}\\
\frac{M^{sc}_K (T)}{M^{sc}_K (0)}\sim \left[\frac{\chi_P^K(0)}{\chi_P^K(T)}\right]^{1/2}=\left[\frac{\condl (0)+2\mean{\bar s s}(0)}{\condl (T)+2\mean{\bar s s}(T)}\right]^{1/2}
\label{scalingK}\\
\frac{M^{sc}_{\bar s s} (T)}{M^{sc}_{\bar s s} (0)}\sim \left[\frac{\chi_P^{\bar s s} (0)}{\chi_P^{\bar s s} (T)}\right]^{1/2}\sim \left[\frac{\mean{\bar s s}(0)}{\mean{\bar s s} (T)}\right]^{1/2},
\label{scalings}
\end{align}
where in \eqref{scalings} we have ignored the anomalous contribution. The above relations can then be directly tested with lattice data, although a proper comparison with the left hand side requires to take into account  finite-size divergences of the kind $\mean{\bar\psi_i \psi_i}\propto m_i/a^2$. This is a known feature of lattice calculation which is avoided by defining properly subtracted condensates.   In figure~\ref{fig-screening} we check this prediction based on WI with subtracted condensates $\Delta_{l,K,s}$ defined in \cite{Nicola:2016jlj} following the lattice prescriptions in \cite{Bazavov:2011nk} with two fitted parameters, and corresponding to the r.h.s of equations \eqref{scalingpi}, \eqref{scalingK}, \eqref{scalings}. The lattice data for screening masses and condensates in that figure are taken from the same collaboration with the same lattice action and size, namely 
\cite{Cheng:2010fe} for the screening masses and \cite{Cheng:2007jq} for the condensates. 

The above results show an excellent agreement (discrepancy below 5\%) between the data and the predicted scaling law, which is remarkable given the different uncertainties involved. Furthermore, the WI explain the sudden growth of  $M_\pi^{sc}$ near the transition, since it is driven by the inverse squared root of the light condensate, while in the $K$ channel the growth is softer due to the presence of the $\conds$ component, and it is even softer for $\bar s s$ for which there is no light condensate contribution.
\section{Saturated Scalar susceptibility: role of the thermal $f_0(500)$}
\label{unit}

As discussed above, susceptibilities are expected to be saturated by inverse squared masses of the corresponding correlators. In the case of the scalar susceptibility, the relevant contributing state should be the $f_0(500)$ for the two-flavour case. That state can be generated within the ChPT formalism by the so called unitarization methods, constructed by demanding unitarity on the partial waves of pion scattering. We  make use  of the so called Inverse Amplitude Method (IAM) extended at finite $T$ \cite{Dobado:2002xf,GomezNicola:2002tn}. One starts from the ChPT series for a  partial wave with given isospin $I$ and angular momentum $J$, $t(s;T)=t_2(s)+t_4(s;T)+\cdots$, where $s$ is the Mandelstam variable, $t_2$ is the tree level from the lowest order ChPT lagrangian ${\cal L}_2$ and $t_4$ includes loop corrections from ${\cal L}_2$ plus tree level contributions from the NLO ${\cal L}_4$, hence absorbing the loop divergences. Such perturbative amplitude satisfies the perturbative unitarity relation $\im t_4 (s+i\epsilon;T)=\sigma_T(s) t_2(s)^2+\cdots$ for $s>4M_\pi^2$ with $\sigma_T (s)=\sqrt{1-4M_\pi^2/s}\left[1+2n_B(\sqrt{s}/2;T)\right]$ and $n_B(x;T)=\left[\exp(x/T)-1\right]^{-1}$ the Bose-Einstein distribution function.   Here, $\sigma_T$ is the thermal two-particle phase space, enhanced with respect to the $T=0$ one by the difference between  emission and absorption scattering processes allowed in the thermal bath \cite{GomezNicola:2002tn}. The IAM amplitude $t^{IAM}$ is obtained by demanding exact thermal unitarity, namely,  $\im t^{IAM}(s+i\epsilon;T)=\sigma_T(s) t^{IAM}(s)^2$ while complying with the low energy expansion, i.e, $t^{IAM}=t_2+t_4+\cdots$. These two conditions lead to:
\begin{equation}
t^{IAM}(s;T)=\frac{t_2(s)^2}{t_2(s)-t_4(s;T)}.
\label{iam}
\end{equation}

Recall that in this construction, thermal unitarity is imposed for the unitarized amplitude. While at $T=0$ this is just the standard unitarity requirement, at finite $T$ requires to promote the  ChPT perturbative relation based on the above mentioned thermal bath processes. As we will see in Sect.~\ref{largeN}, that thermal relation actually holds exactly to leading order in the large-$N$ (number of Goldstone Bosons) framework, which supports the previous assumptions.  In addition, the thermal IAM unitarized amplitude  is analytical in the complex $s$ plane off the real axis, which allows to define the second Riemann sheet amplitude in the usual way, demanding continuity across the unitarity cut. When doing so, resonances in different channels appear as $T$-dependent poles of the amplitude in the second Riemann sheet at $s_p (T)=\left[M_p(T)-i\Gamma_p(T)/2\right]^2$. Thus, for the case of pion scattering, the thermal $f_0(500)$ and $\rho(770)$ are generated \cite{Dobado:2002xf}, their $T=0$ values for the resonance parameters being in good agreement with the PDG ones for  phenomenologically meaningful values of the low-energy constants involved. 
\begin{figure}[h]
\centering
\includegraphics[width=7cm,clip]{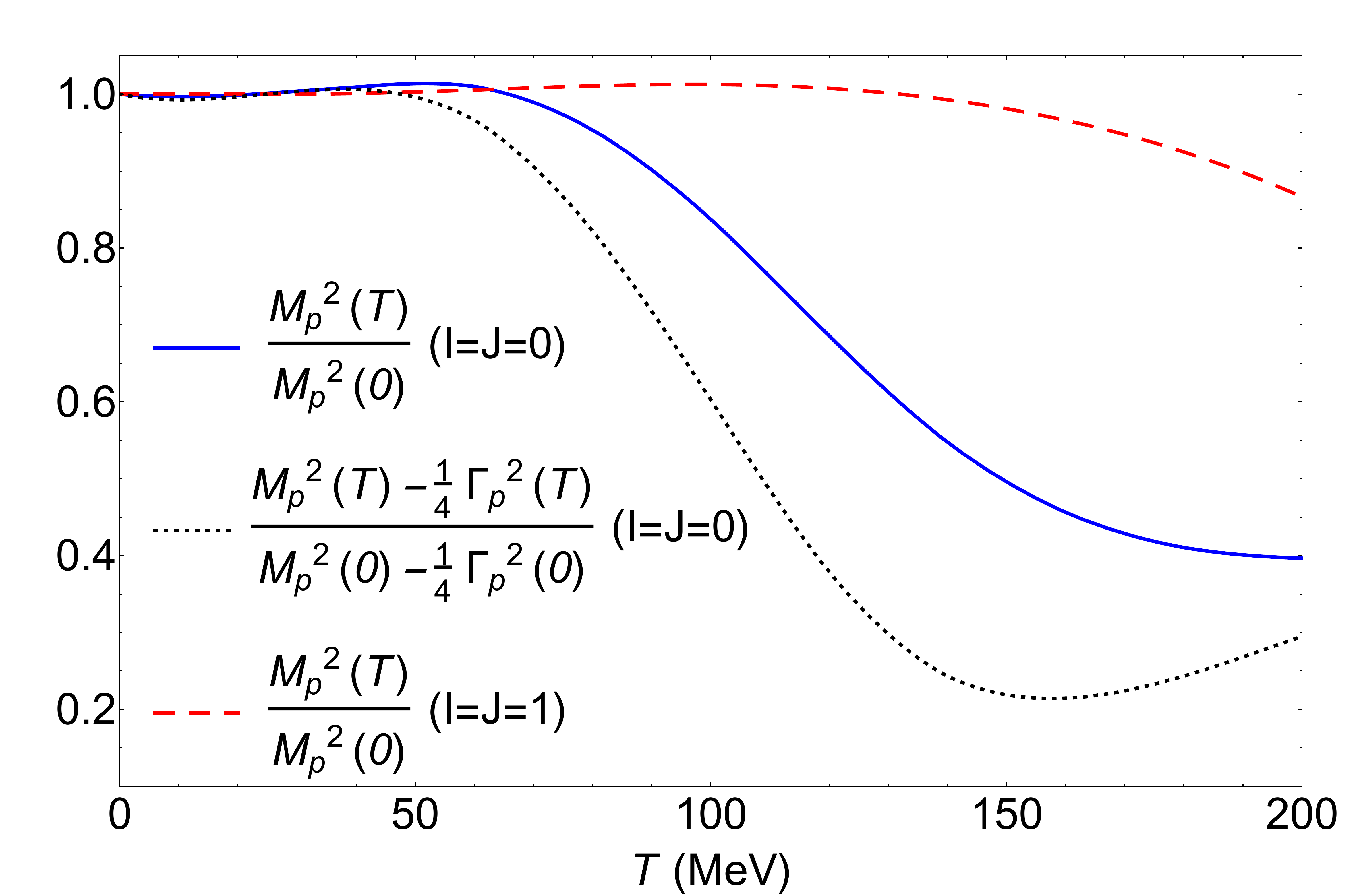}
\includegraphics[width=7cm,clip]{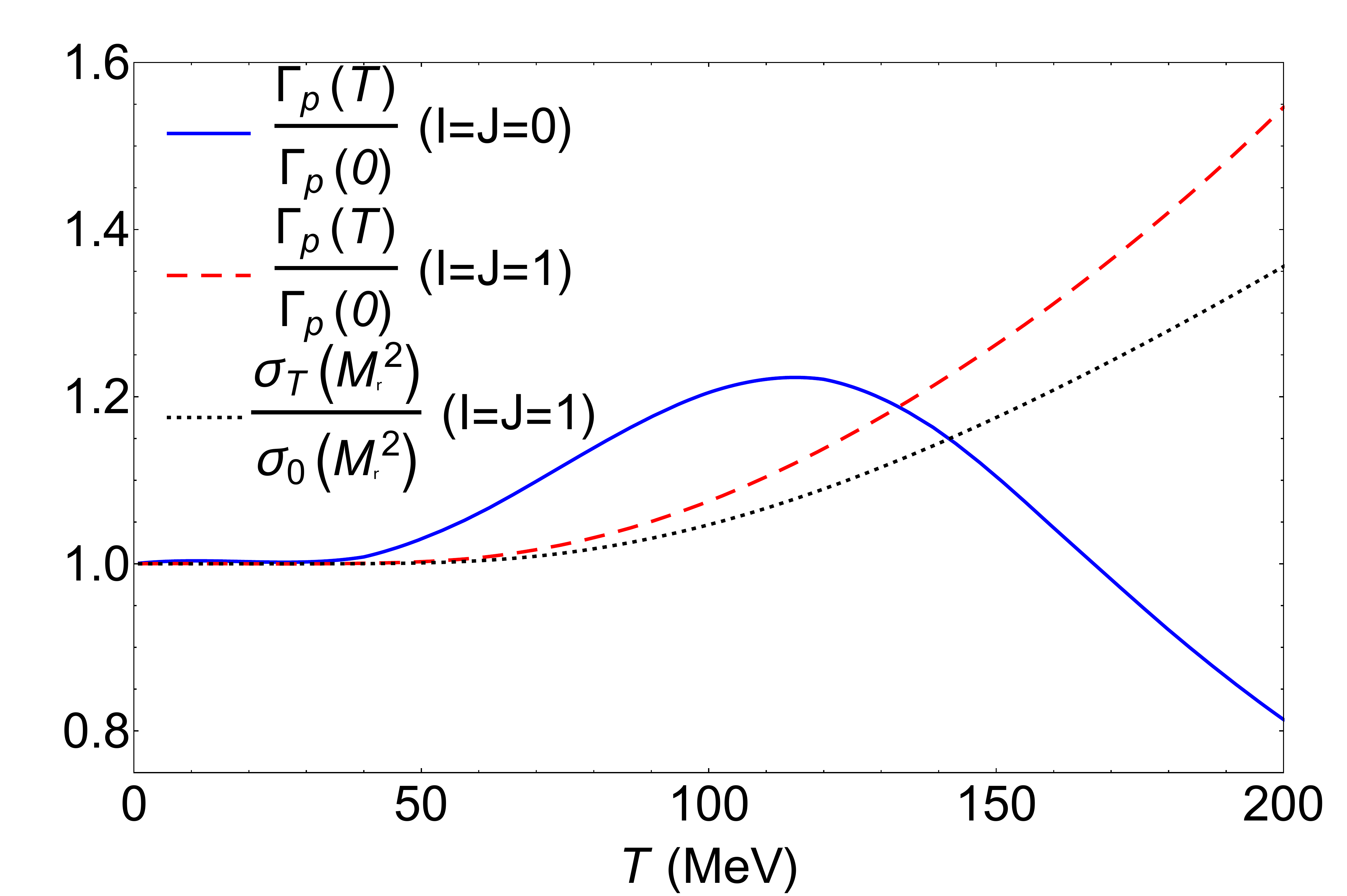}
\caption{Temperature dependence of the $f_0(500)$ ($I=J=0$) and $\rho(770)$ ($I=J=1$) poles}
\label{fig-thermalpoles}       
\end{figure}

\begin{figure}[h]
\centering
\includegraphics[width=7cm,clip]{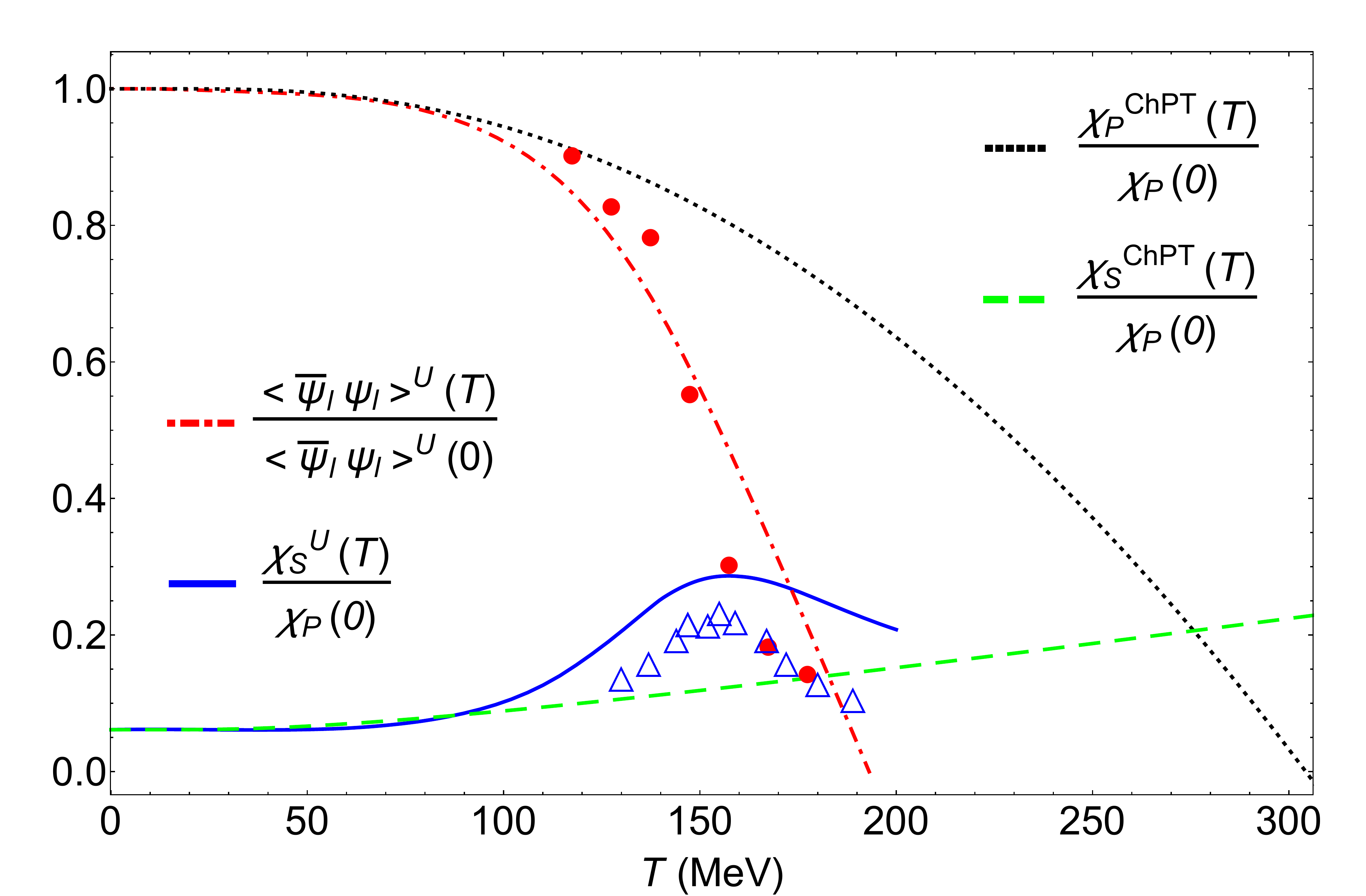}
\caption{Unitarized scalar susceptibility and quark condensate compared to lattice data and ChPT results. The lattice points correspond to figure \ref{fig-scalarpseudolatt} (right).}
\label{fig-unit}       
\end{figure}

In figure~\ref{fig-thermalpoles} we show the temperature dependence of those resonance parameters. The $f_0(500)$ state shows  distinctive features with respect to to the vector channel, which are actually signals of chiral restoration crucial for the present discussion. Thus,  $M_p$ drops with temperature while $\Gamma_p$ increases at low temperatures (driven by phase space thermal enhancement) but decreases closer to the transition. This is mostly due to the phase space reduction induced by the decreasing $M_p$. Interestingly,   $\re s_p (T)=M_p^2 (T)-\Gamma_p^2(T)/4\equiv M_S^2(T)$, which would correspond to the  self-energy real part of a scalar particle exchanged between the incoming and outgoing pions, decreases even faster than $M_p$ thus showing the expected chiral-restoring dropping mass behaviour for this state. Furthermore, following the previous arguments, one can define a saturated unitarized susceptibility as \cite{Nicola:2013vma}:
\begin{equation}
\chi_S^U(T)=\frac{\chi^{ChPT}_S (0) M_S^2(0)}{M_S^2(T)},
\label{susunit}
\end{equation}
where we have normalized to the $T=0$ ChPT model-independent value. This normalization compensates partly the difference between the self-energy real part at $p=0$ (where the susceptibility is defined) and that at the pole position. Hence, with this definition, the minimum observed for $M_S^2(T)$ turns into the expected maximum of $\chi_S$, which turns out to be very close to the predicted lattice value, as showed in figure~\ref{fig-unit}. In that figure we also show a unitarized light condensate defined by integration in mass of $\chi_S^U$, assuming the same $T/M$ dependences of the condensate and the susceptibility  as in  perturbative ChPT \cite{Nicola:2013vma}. 
The results in figure~\ref{fig-unit} show a clear improvement of the unitarized susceptibility to describe lattice data, which highlights the  importance of including properly the thermal $f_0(500)$ state.  Recall that the above result is not fitted to lattice points but it is just the prediction of the light  sector with parameters chosen to describe $T=0$ resonances. Further improvement can be achieved by incorporating higher mass hadron states and fitting the lattice data. Note also that $\chi_S^U$ remains close to the ChPT result for low temperatures, even though they come from a complete different description and are matched only at $T=0$. In addition, the unitarized condensate intersects $\chi_S^U$ close to its maximum, in agreement with the expectations of $\pi-\sigma$ partner degeneration discussed above. 
\section{Large $N_{GB}$ approach}
\label{largeN}

In this section we will show how some of the  ideas developed above are actually realized analytically in a particular framework within the effective lagrangian approach, namely, the large-$N$ one where $N$ is the number of Goldstone Bosons. That scheme provides actually a resummation procedure for infinite subsets of ChPT diagrams and is based on the $S^N=O(N+1)/O(N)$ formulation of the non-linear sigma model, i.e, the lowest order ChPT lagrangian.   This is the most general  $O(N+1)$-invariant (in the chiral limit) and $S^{N}$ covariant lagrangian to lowest order in derivatives. Thus, from the lagrangian vertices, different observables can be calculated identifying the dominant diagrams, including finite temperature corrections within the standard imaginary-time formalism. This program has been carried out recently in the chiral limit for pion scattering, the thermal $f_0(500)$ and the saturated susceptibility in \cite{Cortes:2015emo} and for the quark condensate and scalar susceptibility derived from the partition function in \cite{Cortes:2016ecy}.  Higher order lagrangians absorb loop divergences and it can be shown that a $T=0$ vertex renormalization guarantees the finiteness of the results (see details in \cite{Cortes:2015emo}). 

\begin{figure}[h]
\centering
\includegraphics[width=8cm,clip]{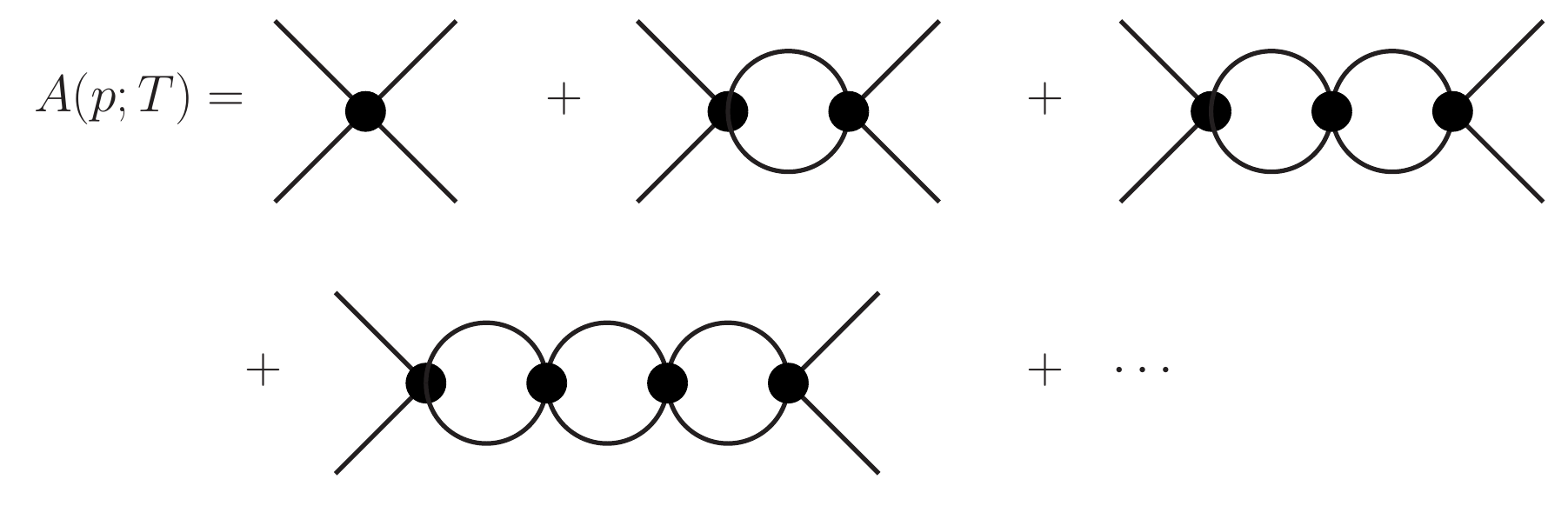}
\caption{Diagrams contributing to the leading order large-$N$ NGB scattering amplitude}
\label{fig-largenamp}       
\end{figure}

\begin{figure}[h]
\centering
\includegraphics[width=8cm,clip]{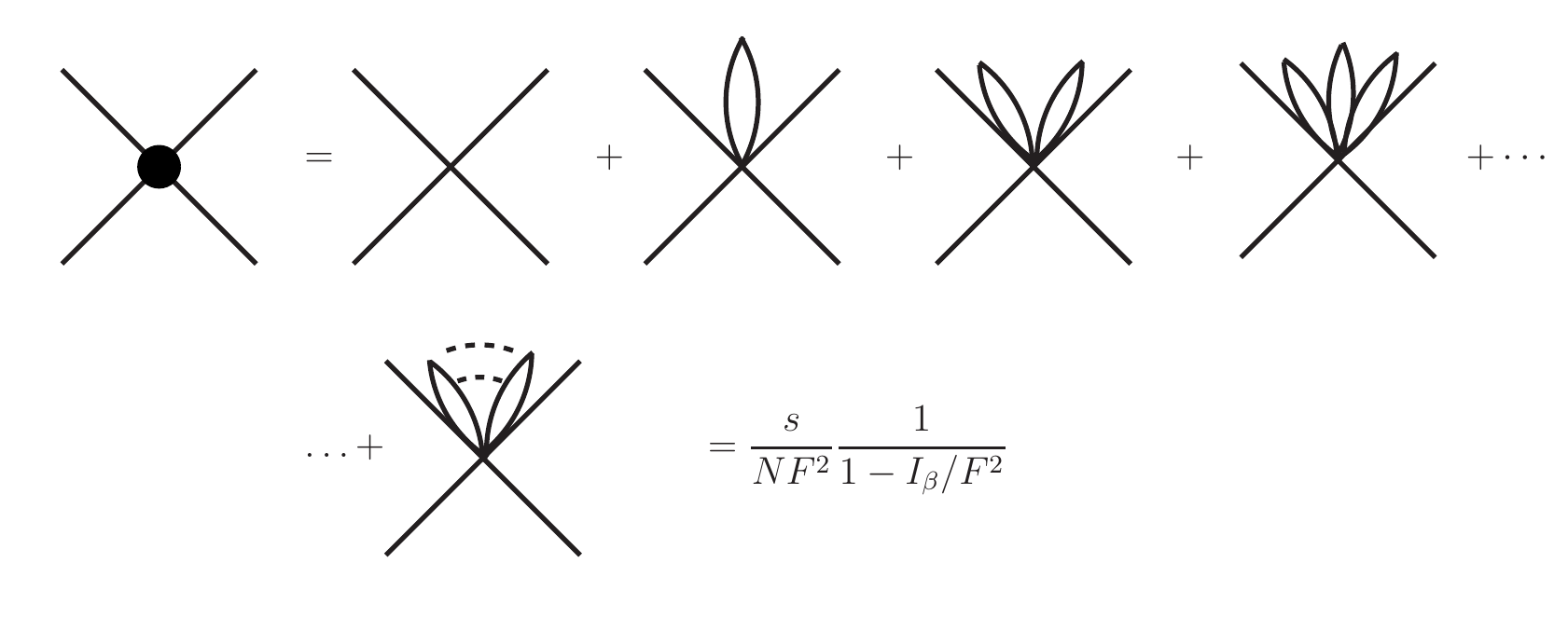}
\caption{Thermal effective vertex entering the large-$N$ scattering amplitude}
\label{fig-largenampvert}       
\end{figure}

For the analysis of pion scattering, the relevant dominant diagrams are showed in figure~\ref{fig-largenamp}, in terms of the effective thermal vertex defined in figure~\ref{fig-largenampvert}, which resums  tadpole insertions and is  a genuine thermal correction. Here, $A(p;T)$ is the scattering function in terms of which all NGB scattering processes are defined and we denote $I_\beta=T^2/12$, the thermal tadpole correction to $G_1\equiv G(x=0)$, with $G$ the NGB  propagator, and $NF^2$ is the pion decay constant squared. 

\begin{figure}[h]
\centering
\includegraphics[width=7cm,clip]{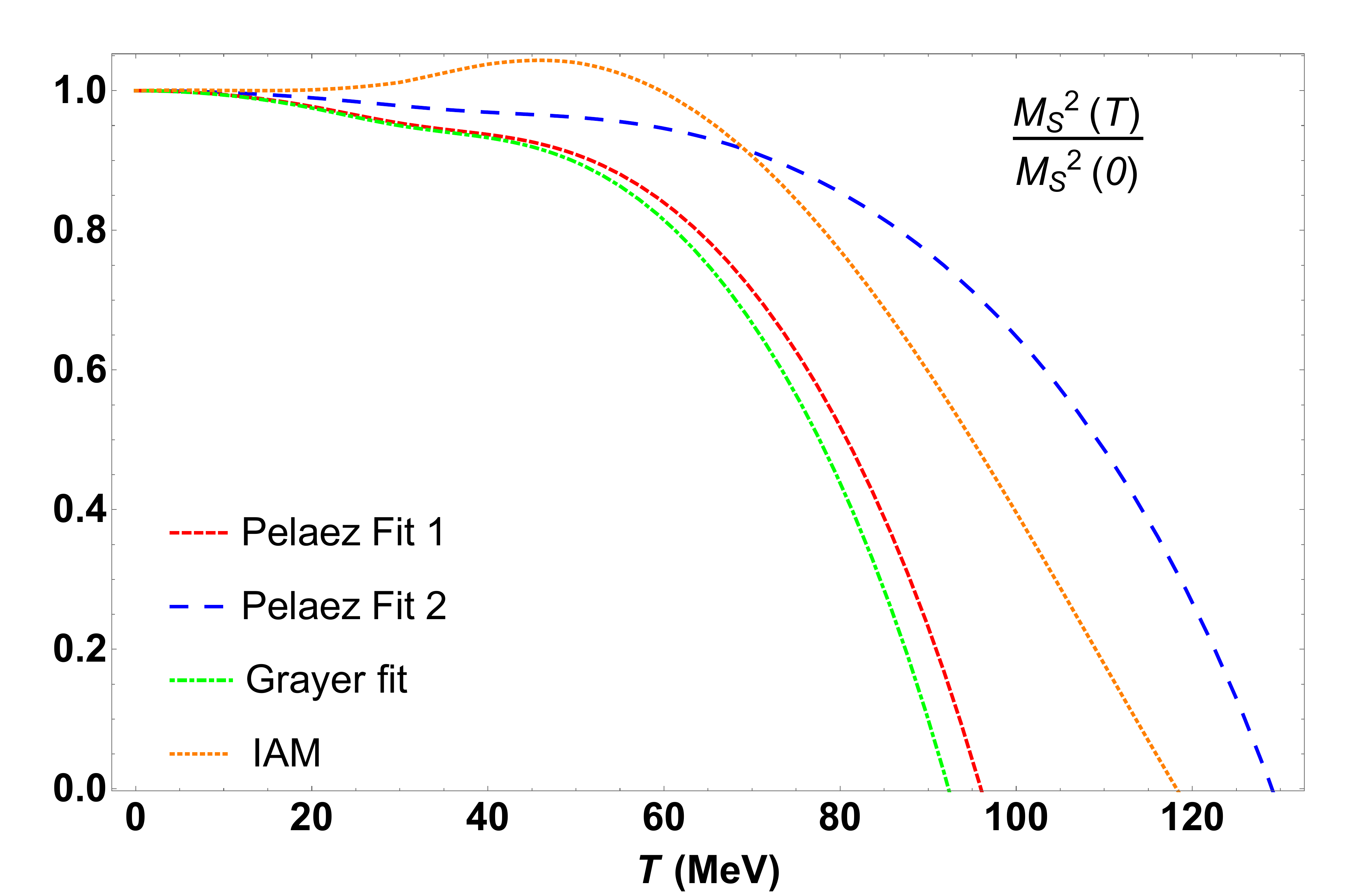}
\includegraphics[width=7cm,clip]{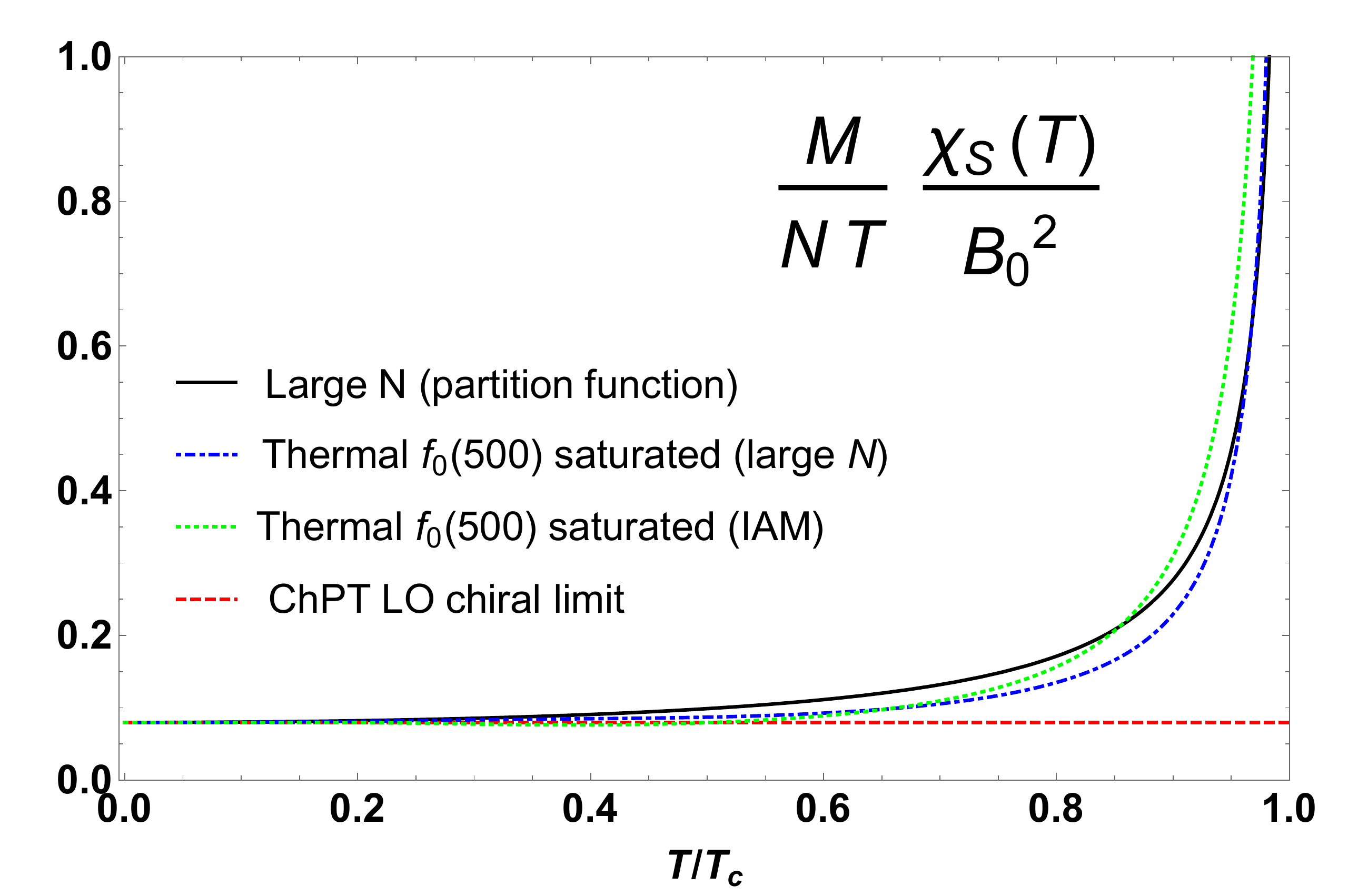}
\caption{Left:Scalar mass for the $f_0(500)$ resonance within the large-$N$ approach for different fits as discussed in \cite{Cortes:2015emo}. Right: Comparison of different approaches for the scalar susceptibility in the chiral limit}
\label{fig-Mslargen}       
\end{figure}

Once the amplitude is renormalized, it depends only on two parameters, $F$ and the renormalization scale $\mu$ at which the low-energy renormalizing constants are chosen to have vanishing finite part. Those parameters are then fixed to low-energy scattering phase shift for the $I=J=0$ partial wave, which is the dominant one at large $N$. Even though the data correspond to nonzero pion mass, the fits are rather good \cite{Cortes:2015emo}, giving a larger value for $F$ than expected precisely to absorb those finite mass effects.  An important result in this analysis is that the scattering amplitude, including finite $T$ corrections, satisfies automatically  the thermal unitarity relation $\im t=\sigma_T \vert t\vert^2$ in the $I=J=0$ channel and has the expected analytical behavior. This allows to generate the $f_0(500)$ state also in this approach, giving rise   to $T=0$  values of $M_p$ and $\Gamma_p$ in reasonable agreement with the PDG ones for the above mentioned parameter fits. At nonzero $T$, the evolution of $M_S^2(T)$  defined in Sect.~\ref{unit}, which would saturate the scalar susceptibility, is showed in figure~\ref{fig-Mslargen} (left) for differents fits to scattering data \cite{Cortes:2015emo}. Since we are working in the chiral limit, the susceptibility is expected to diverge at $T_c$ within a second-order phase transition regime and so it does, with  $M_S^2(T)$ vanishing  at the $T_c$ values given in the figure.  They range from 90-130 MeV, being not far  from the expectations of lattice results in the chiral limit. The critical exponents for $\chi_S$ are also in agreement with lattice, thus providing a consistent picture within this framework \cite{Cortes:2015emo}.

\begin{figure}[h]
\centering
\includegraphics[width=7cm,clip]{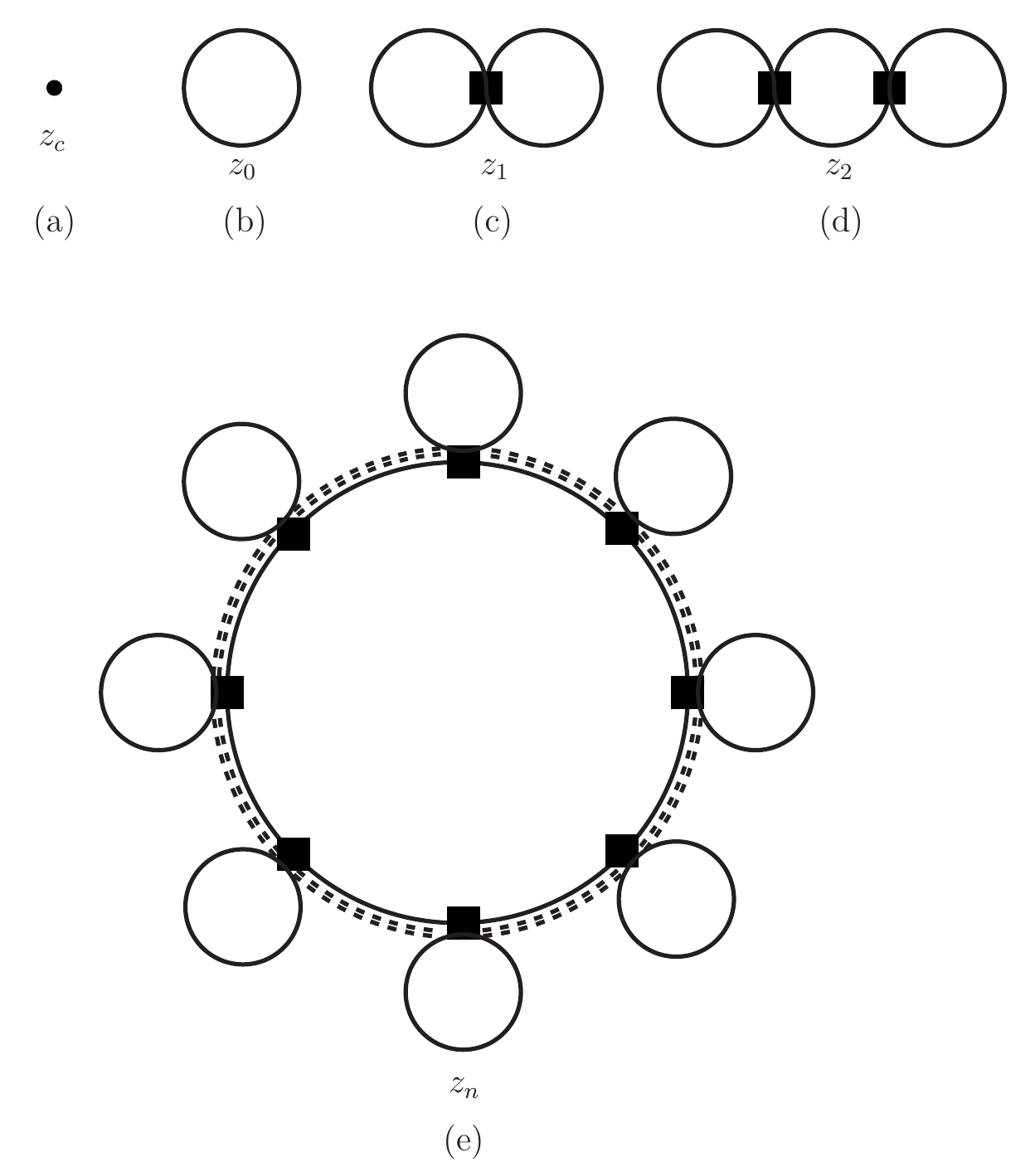}
\caption{Diagrams contributing to leading order in $N$ and to $\Od(M^3)$ to the partition function}
\label{fig-parfundiag}       
\end{figure}

An alternative approach within the large-$N$ framework is to work out the diagrammatic expansion of the partition function for finite NGB mass $M$, in order to extract the quark condensate and the scalar susceptibility as first and second derivatives with respect to $M$ of the free energy, respectively \cite{Cortes:2016ecy}. The leading diagrams in this case are displayed in figure~\ref{fig-parfundiag} for the chiral limit calculation, with the effective mass vertex in figure~\ref{fig-parfundiag}, where $g=-\frac{8}{x^2}\left[\sqrt{1-x}-1+\frac{x}{2}\right]$. In figure~\ref{fig-Mslargen} (right) we show the different approaches to the scalar  susceptibility, incluing the IAM described in the previous section, in terms of $T/T_c$ with different $T_c$ for every approach, so that we can compare their critical behaviour in the chiral limit. In that plot, we have defined the saturated susceptibility in the chiral limit as $\displaystyle\chi_S=\frac{NTB_0^2}{4\pi M}\frac{M_S^2(0)}{M_S^2(T)}$ so that it reproduces the expected ChPT behaviour at low $T$ (see Sect.~\ref{part}). When comparing the thermal evolution and critical exponents of the different approaches, they are consistent within the uncertainties expected for the large-$N$ approach.  
\begin{figure}[h]
\centering
\includegraphics[width=8cm,clip]{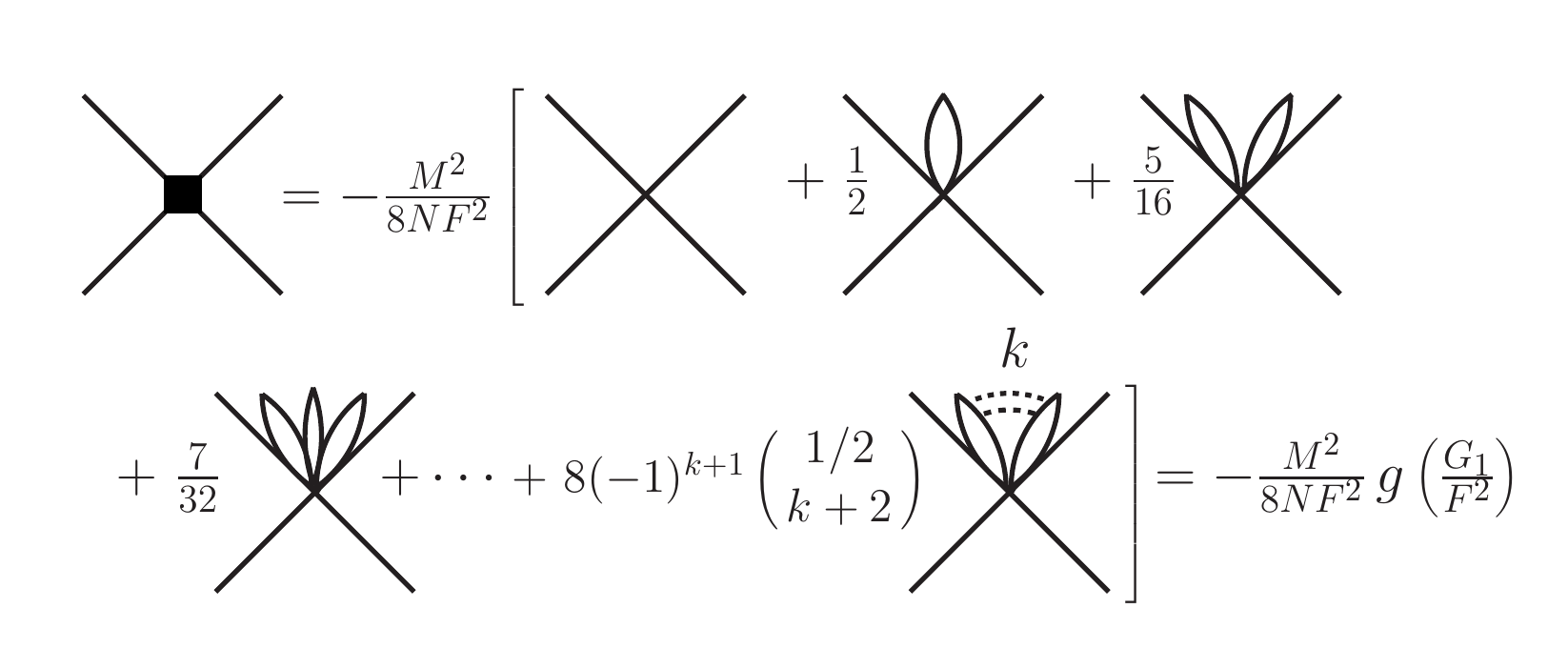}
\caption{Effective mass thermal vertex entering the partition function analysis at large $N$.}
\label{fig-ma}       
\end{figure}
\section{Conclusions}
\label{conclusions}

We have presented  recent results regarding chiral symmetry restoration in QCD.  Our theoretical approach, based on the effective theories, incorporates ideas from QCD Ward Identities, unitarization and the large-$N$ framework. WI relating quark condensates and pseudoscalar susceptibilities provide a way to understand the thermal behaviour of lattice screening masses near the transition. On the other hand, the thermal $f_0(500)$ pole generated from unitarized schemes plays a crucial role in the description of the scalar susceptibility, yielding a peak compatible with lattice data and a  picture consistent with $O(4)$ degeneration. The large-$N$ framework allows to verify thermal unitarity and to describe a second-order phase transition in the chiral limit consistent with lattice results and other theoretical analysis. Within these ideas, the analysis of the possible restoration of $U_A(1)$ at the chiral transition within the context of the scalar-pseudoscalar nonet degeneration is a matter for future work. 
\section*{Acknowledgments}
Work partially supported by the Spanish Research contracts FPA2014-53375-C2-2-P and FIS2014-57026-REDT. J.R.E. gratefully acknowledges financial support by the DFG (SFB/TR 16, ``Subnuclear Structure of Matter'') and  the Swiss National Science Foundation. S.C.  thanks Prof. J.R. Rold\'an, the High Energy Physics group of Universidad de los Andes and COLCIENCIAS for  financial support. 
%
%
%
%

\end{document}